\documentclass[preprint,review,12pt]{elsarticle}
\usepackage[utf8]{inputenc} 
\usepackage[T1]{fontenc}    
\usepackage{amsfonts}       
\usepackage{amsmath}
\usepackage[thinc]{esdiff}

\newcommand{\icol}[1]{
  \left(\begin{smallmatrix}#1\end{smallmatrix}\right)%
}

\usepackage{amssymb}

\usepackage{lineno}
\begin{document}
\begin{frontmatter}



\title{Social dilemmas, network reciprocity, and small-world property}
\author[aff1]{F.B. Pereira}
\author[aff1]{R.S. Ferreira\corref{cor1}}
\author[aff2]{D.S.M. Alencar}
\author[aff2]{T.F.A. Alves}
\author[aff3]{G.A. Alves}
\author[aff2]{F.W.S. Lima}
\author[aff3]{A. Macedo-Filho}
\cortext[cor1]{ronan.ferreira@ufop.edu.br}
\affiliation[aff1]{organization = {Departamento de Ciências Exatas e Aplicadas},
            addressline         = {Universidade Federal de Ouro Preto},
            city                = {João Monlevade},
            postcode            = {35931-008},
            state               = {Minas Gerais},
            country             = {Brazil}}
            
\affiliation[aff2]{organization={Departamento de Física},
            addressline={Universidade Federal do Piauí},
            city={Teresina},
            postcode={57072-970},
            state={Piauí},
            country={Brazil}}
            
\affiliation[aff3]{organization = {Departamento de Física},
            addressline         = {Universidade Estadual do Piauí},
            city                = {Teresina},
            postcode            = {64002-150},
            state               = {Piauí},
            country             = {Brazil}}

\begin{abstract}
We revisit two evolutionary game theory models, namely the Prisoner and the Snowdrift dilemmas, on top of small-world networks. These dynamics on networked populations (individuals occupying nodes of a graph) are mainly concerned with the competition between cooperating or defecting by allowing some process of revision of strategies. Cooperators avoid defectors by forming clusters in a process known as \emph{network reciprocity}. This defense strategy is based on the fact that any individual interacts only with its nearest neighbors. The minimum cluster, in turn, is formed by a set of three completely connected nodes, and the bulk of these triplets is associated with the transitivity property of a network. We show that the transitivity increases eventually, assuming a constant behavior when observed as a function of the number of contacts an individual has. We investigate the influence of the network reciprocity on that transitivity-increasing regime on promoting cooperative behavior. The dynamics of small-world networks are compared with those of random regular and annealed networks, the latter typically studied as the well-mixed approach. The Snowdrift Game converges to an annealed scenario as randomness and coordination numbers increase. In contrast, the Prisoner's Dilemma becomes more severe against the cooperative behavior under an increasing network reciprocity regime.
\end{abstract}



\begin{keyword}
Evolutionary games \sep Complex networks \sep Dynamical processes



\end{keyword}

\end{frontmatter}







\section{Introduction}
\label{intro}
How can cooperative behavior emerge and persist when alternative strategies are spatially distributed within a population of selfish individuals? Similar questions have been addressed in many contexts as diverse as biological or social dilemmas \cite{hofbauer2003evolutionary,axelrod1981evolution,turner1999prisoner,szabo2007evolutionary, santos2014biased}. Evolutionary dynamics have been characterized in numerous scenarios by their application to structured populations. One of its contributions can be highlighted by defining a framework for studying strategic decision-making and the mathematical modeling of conflict and cooperation among greedy individuals. The concern in evolutionary game theory is that the success of a strategy is not just determined by how good the strategy is in itself; instead, it is a question of how good the strategy is in the presence of other alternatives and of the way the distribution of strategies take place within a population. Evolutionary games combined with network theory may explore structural and dynamical mechanisms for promoting cooperative behavior.

Under the evolutionary version of a classical two-player game, each player can opt for one of the two strategies, \emph{cooperate} ($C$) or \emph{defect} ($D$). At the end of each combat, both players receive a payoff. The player’s payoff is a quantity that depends on the choice strategy and is determined by mapping into a payoff matrix, $\mathbf{M}$. Depending on the balance between \emph{benefit} ($b$) and \emph{cost} ($c$), thus are established a scale to explore dynamics interactions between individuals. Each player receives the \emph{reward} $R$ on mutual cooperation, whereas reciprocal defections return the \emph{punishment} $P$. On changed strategies, a cooperator receives the \emph{suckers} payoff $S$, and the defector wins a \emph{temptation} (to defect) reward $T$. These game parameters $R$, $P$, $S$, and $T$ are the entries of the payoff matrix. What makes the difference between the two games studied in this work is the scale assumed of the parameters, hence the balance benefit \emph{versus} cost.

On one hand, Prisoner Dilemma game (PD), the payoff's scale is defined as $T > R > P > S$, where $T=b$, $R=b-c$, $P=0$, and $S=-c$. So, the dilemma arises since the best choice is when defecting, but if both players choose to defect in combat, they lose more since providing the highest total income for mutual cooperation. On the other hand, when collaborative work takes place by sharing costs of cooperation, the situation is illustrated in the Snowdrift game (SD). Its payoff's scale is defined as $T > R > S > P$, where $T=b$, $R=b-c/2$, $S=b-c$, and $P=0$. As defecting is not reduced by the cost, it remains favored even if the opponent player cooperates. So, the best choice is cooperative choice, even among selfish individuals. Since these games have no memory based on past confrontations, topological arrangements strongly influence cooperative behavior.

When these games occur on network topologies, each individual is represented by nodes and can play against their nearest neighbors assigned by links, following some connectivity distribution. Since each individual assumes a strategy, the dynamic is due to some process allowing a revision of strategies. Here, we investigate an update rule based on the \emph{suggestion process} where an individual can accept the same strategy one of its neighbors suggested. A possible mechanism to promote cooperative behavior is the so-called \emph{network reciprocity} \cite{nowak1992evolutionary,hauert2004spatial,nowak2006five,roca2009effect}. It can be thought of as a defense game plan to assure the survival of a strategy and is based on two facts: (\emph{i}) Individuals interact only with their first neighborhood; (\emph{ii}) if all individuals in the same neighborhood support a selfsame strategy, this clustering does not change their choices.

This defense mechanism can be particularly explored in small-world structures. This network class has the main feature of combining a high clustering with a low shortest path between pairs of nodes. Whereas high clustering is typically found on regular structures, a low average shortest path is commonly observed on random networks. It is possible to use link rewiring, which introduces shortcuts that connect different network parts. This process leads to a shortening of the path between any pair of nodes and a decrease in the aggregation of nodes in the network. Equivalently, it turns up by reformulating triangles previously established among the three nearest nodes, which, in turn, are the minimum clustering arrangement. The bulk of these triangles is associated with the transitivity property of a network; the transitivity increases, eventually assuming a constant behavior when observed as a function of the number $k$ of contacts of an individual.

The role of triangles over the PD's dynamics was specially studied on a one-dimensional lattice where individuals are connected to $k = 4$ neighbors by Vukov \textit{et al.} - Ref.~\cite{PhysRevE.77.026109}. Also influenced by this work, it has reported the impact on evolutionary games given multiplex scale-free networks as well as clustered scale-free networks~\cite{PhysRevE.89.052813,PhysRevE.82.047101} and others strategies update, mobility, and group interaction~\cite{PhysRevE.78.066101,chen2008interaction,PhysRevE.80.056112,Perc_2011,JIA2024115333,Li_2024,A_li2016,PhysRevE.79.067101,Wu2017,Gang2008}. In the present work, in turn, we are going to show that there is an increasing regime for network reciprocity, for low values of $k$ (including $k=4$) where the transitivity property, as a function of $k$, deviates quantitatively from its analytical expectation, for small-world networks. Eventually, this function reaches a constant behavior, corresponding to the analytical approach for higher values of $k$. Therefore, the present work is concerned with the dynamical behavior of cooperation in this increasing regime of network reciprocity when PD and SG games occur on small-world topologies.

\section{Methods}
\label{metho}
\subsection{Evolutionary games}
The models PD and SG were implemented following the standard two-player scheme: Each player assumes an unconditional strategy, $C$ (cooperator) or $D$ (defector). To characterize the dilemma, we used the constraint $2R>T+P$ to the payoff scale, implying that temptation and sucker rewards can assume $T=[R,2R]$ and $S=[R,P]$. These ranges were represented by the payoff matrix $\mathbf{M}$, in which the elements or entries $M_{mn}$ represent the payoff earned by the row player:
%
\begin{equation}
\bordermatrix{     
          & \mathbf{C}  & \mathbf{D}      \cr
    \mathbf{C}     & R  & S     \cr
    \mathbf{D}     & T  & P      \cr
}
\end{equation}
For example, when the row player decides on cooperation (first line), and the column player opts for defection (second column), the row player earns the \emph{Sucker} payoff $M_{12}=S$. On the other hand, if the row player decides to defect (second line) and meets a collaborator (first column), the defector player earns the \emph{Temptation} payoff $M_{21}=T$. When both players assume the same strategy, $C$$C$ or $D$$D$, the row player receives the payoff $M_{11}=R$ or $M_{22}=P$, respectively.
By setting $R=1$ and $P=0$, we can set free $T$ and $S$ parameters. So, the matrix payoff for PD dynamics is
   \begin{equation}
       \mathbf{M}_{PD}=
       \begin{pmatrix}
           1 & -S\\
           T & 0
       \end{pmatrix}
       \label{mtr:PD}
   \end{equation}
and for SG yields
   \begin{equation}
       \mathbf{M}_{SG}=
       \begin{pmatrix}
           1 & S\\
           T & 0
       \end{pmatrix}
       \label{mtr:SG}
   \end{equation}

\subsubsection{Spatial version and the update strategy rule}
In the spatial version of this evolutionary dynamics, the strategy $u_i=C$ or $u_i=D$ of each node (player) $i$ is represented by a 2-dimensional vector space of orthonormal bases, corresponding to $C=\icol{1\\0}$ and $D=\icol{0\\1}$, and the payoff of each node is given by the inner product
\begin{equation}
f_i=\sum_{j \in \Omega_i}u^T_i\mathbf{M}u_j
\end{equation}
where the summation runs over all nodes $j$ connected to the node $i$ ($\Omega_i$ represents the neighborhood of the node $i$) and $\mathbf{M}$ is given by equation (\ref{mtr:PD}) or (\ref{mtr:SG}), depending on the game dynamics.

From a total of $n$ players, if $n_c$ of them choose to cooperate, then $n_d$ ($=n-n_c$) players assume the defecting strategy. 
The rate equation for the density of cooperators $\rho_c=n_c/n$ concerning the time $t$, based on the principle of replicating dynamic equations \cite{taylor1978evolutionary}, is given by 
\begin{equation}
	\diff{\rho_c(t)}{t} = \rho_c(t)(1-\rho_c(t))\left[f_c(\rho_c)-f_d(\rho_c)\right]
\end{equation}
where
\begin{equation} \label{eq1}
 \begin{split}
  f_c(\rho_c) & = \rho_c(t)R+(1-\rho_c(t))S \\
  f_d(\rho_c) & = \rho_c(t)T+(1-\rho_c(t))P
 \end{split}
\end{equation}
are the payoffs of a cooperator and a defector, respectively.

In simulations, size systems were used from $N=10^4$ to $N=10^6$ nodes (players), performed as follows: At each round, a player $x$ fights against their $k$ neighbors, and its payoff is computed. Then, a player $y$ from the player's neighborhood $x$ is chosen by chance. If its strategy differs from $x$, its payoff is computed. With probability $W$ player $x$ assume the strategy of player $y$, where
\begin{equation}
 W(u_x \leftarrow{u_y})=\frac{1}{1 + \exp[-\beta(f_y - f_x)]}
\end{equation}
and $\beta$ controls perturbations on the revision process of strategies. For successful strategy adoption, both the strategy and the payoff are updated following an asynchronous protocol.

After a relaxation time $t_r$, the dynamics reach a stationary regime, for which $d\rho_c/dt=0$. During this regime, the average density $\rho_c$ of cooperators was determined on an interval sampling time $t_s\sim t_r$. For the initial state it was used ensembles of the order from $10^1$ to $10^2$ network realizations with $\rho_c(t_0)$ chosen at random ($t_0$: initial time), guided by fluctuations $\delta\rho_c(t)\leq10^{-4}$.

\subsection{Small-world networks: The Watts-Strogatz model}
\label{ssec:NetModels}
%
A small-world network can be built as follows: A number $N$ of nodes are ordered initially in a regular ring lattice, \textit{i.e.}, in a one-dimensional lattice with periodic boundary conditions. Each of them has $k$ connections with the nearest neighbors. One node is chosen randomly, and one of its $k$ links is selected by chance in the clockwise sense. This link is redirected with probability $p$ to another node, uniformly chosen in the network, avoiding duplicate connections. This procedure is repeated clockwise until an entire lap over the one-dimensional ring is complete.

The rewiring rules generate connected networks and conserve the number of links, implying that $\langle k\rangle = k$. This rewiring process introduces $pNk/2$ long-range connections, reducing the average shortest path $\langle\ell\rangle$ between nodes. For some values of $p$, typically found in the range [$10^{-3}$,$10^{-1}$], one observe the small-world phenomena~\cite{watts1998collective}.

Three representative values of the rewiring probability were used: $p=0.0$ (regular), $p=0.1$ (small-world), and $p=1.0$ (random). For $p=0.0$, network is $k-$regular with $\langle\ell\rangle\approx N/2k$ and a high average clustering coefficient $\langle c\rangle \approx 3/5$ \cite{albert2002statistical}. In this case, the network ensemble shows a fixed total number of links $L$ besides the constraint that each node has the same number of connections, $k_i=k$. This way, $L=\sum_{i<j}g_{ij}$ and
    \begin{equation}
        k_i=\sum_{j=1}g_{ij}.
        \label{eq:ki}
    \end{equation}

On the other hand, for $p$ close to $1$, the network converges to both vanishing clustering $\langle c\rangle \sim k/N$ and small average shortest path $\langle\ell\rangle\sim ln(N)/ ln(k)$, typically found on random networks. A broad range of $p$ values occurs where a large $\langle c\rangle$ and a short $\langle\ell\rangle$ are concurrent. However, for any finite $p$ and sufficiently large sizes, shortcuts in the networks render the small-world property \cite{barrat2000properties}. So, for any value of $p>0$, the constraint presented on equation (\ref{eq:ki}) turns up an argument only in average, where $\langle k\rangle=k$.

\subsection{Others network models}
As a guide to compare the results observed on WS networks, the evolutionary dynamics were also studied in addition to canonical network models. For the case of structures based on lattice $k-regular$ fashion, we used networks where each node has a \emph{von Neumann} neighborhood, with $4$ nearest neighbors. On the other hand, to compare the cases where $p>0$, two network models was used, namely:

\subsubsection{Random regular networks} Random regular networks (RRN) restrict the degree of nodes, in which each node $i$ has the same degree $k_i=k$. However, nodes are connected by chance and self, and multiple connections are forbidden. So, the connectivity distribution is $P(k)=\delta_{k,q}$. The RRN has finite length loops only when the network size $N$ is finite and is locally tree-like so that it can be handled as random Bethe lattices \cite{dorogovtsev2008critical,dorogovtsev2010lectures}. This way, it is possible to compare the role of degree-degree correlations with WS-networks with $p=1.0$: For WS-networks, $P(k)$ decays exponentially fast for large $k$, producing heterogeneous degree distributions in a strict sense \cite{Ferreira2013}.

\subsubsection{Annealed networks} Annealed networks (ANN) have a structural modification concerning restoring the entries $a_{ij}$ of the adjacency matrix by a probability that a node $i$ of degree $k_i$ is connected to a node $j$ with $kj$ connections. Structural interactions are modified in a characteristic time lower than the interval $dt$ of the dynamic process. Since all links are redistributed at a rate lower than the dynamical time interval, finite-length loops are broken from one-time step to another. This influence of loops is recognized to be important over critical behavior of dynamical processes \cite{dorogovtsev2008critical,PhysRevE.83.066113}. Therefore, the ANN approach is considered an enhanced mean-field theory suitable for studying networks with heterogeneous connectivity distribution.

\section{Results and discussions}
\subsection{Structural analysis}
\label{sec:Str-anl}
The \emph{transitivity} of a network is the relative bulk of triangles among all connected triplets of nodes \cite{luce1949method}. A \emph{triangle} is a set of three completely connected nodes. On the other hand, a triplet can be either a triangle or a set of three nodes connected by only two links. From this point of view, the transitivity can be interpreted as the probability of finding a connection between two nodes having a common neighbor. Therefore, this measure acts as a clustering coefficient, going from $0$ to $1$. For $p=0$, each node has $2k$ neighbors, yielding in $3k(k-1)/2$ links among them. For values of $p>0$, two neighbors of a node $i$, previously connected at $p = 0$, remain neighbors of $i$ and linked with probability $(1-p)^3$. This probability holds up to order $1/N$, where $N$ is the total number of nodes (network size). Since the average clustering coefficient is defined by $\left[\sum_i\frac{e_i}{k_i(k_i-1)/2}\right]/N$, where $e_i=(1/2)\sum_{jl}a_{ij}a_{jl}a_{li}$ is the number of links between them, and $a_{mn}$ are entries of the adjacency matrix, in which if the nodes $m$ and $n$ are connected $a_{mn}=1$ and $a_{mn}=0$ otherwise. Therefore, one may define this concept corresponding to the transitivity of triplets $\tau$ as a function of rewiring probability $p$,
 \begin{equation}
  \langle\tau(p)\rangle\approx\frac{3(k-1)}{2(2k-1)}(1-p)^3,
  \label{eq.trn}
 \end{equation}
considering additional terms of order of $1/N$ \cite{wasserman1994social,barrat2000properties}. These patterns of interconnections appear in many complex networks, and they are considered a kind of basic building blocks, specifying structural features of a network \cite{milo2002network}.

\begin{figure}[h]
 \centering
 \includegraphics[scale=0.36]{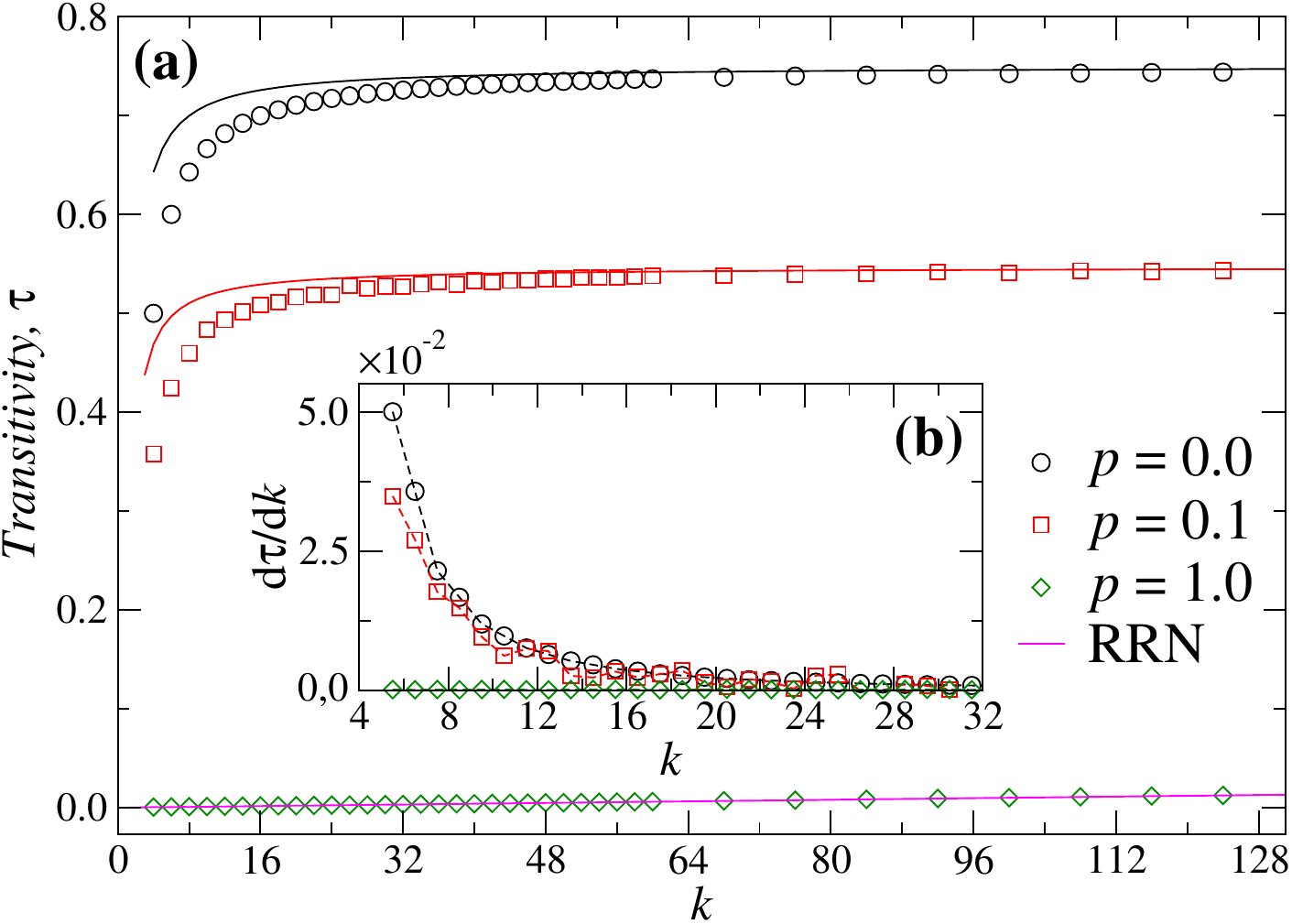}
 \caption{\textbf{\emph{Transitivity}}. Panel $\mathbf{(a)}$: Analytical descriptions offered by equation (\ref{eq.trn}) were compared to analysis from simulations (symbols). For curves $p=0.0$ and  $p=0.1$, a transient region is marked by a decreasing growth rate, shown in panel $\mathbf{(b)}$. It takes place for values up to $k\lesssim24$. For the curve $p=1.0$ and RRN, one expects no corrections on the scale due to the strong presence of shortcuts. It used WS networks with $N=10^4$ nodes.}
 \label{fig01}
\end{figure}

\begin{figure}[h]
 \centering
 \includegraphics[scale=0.36]{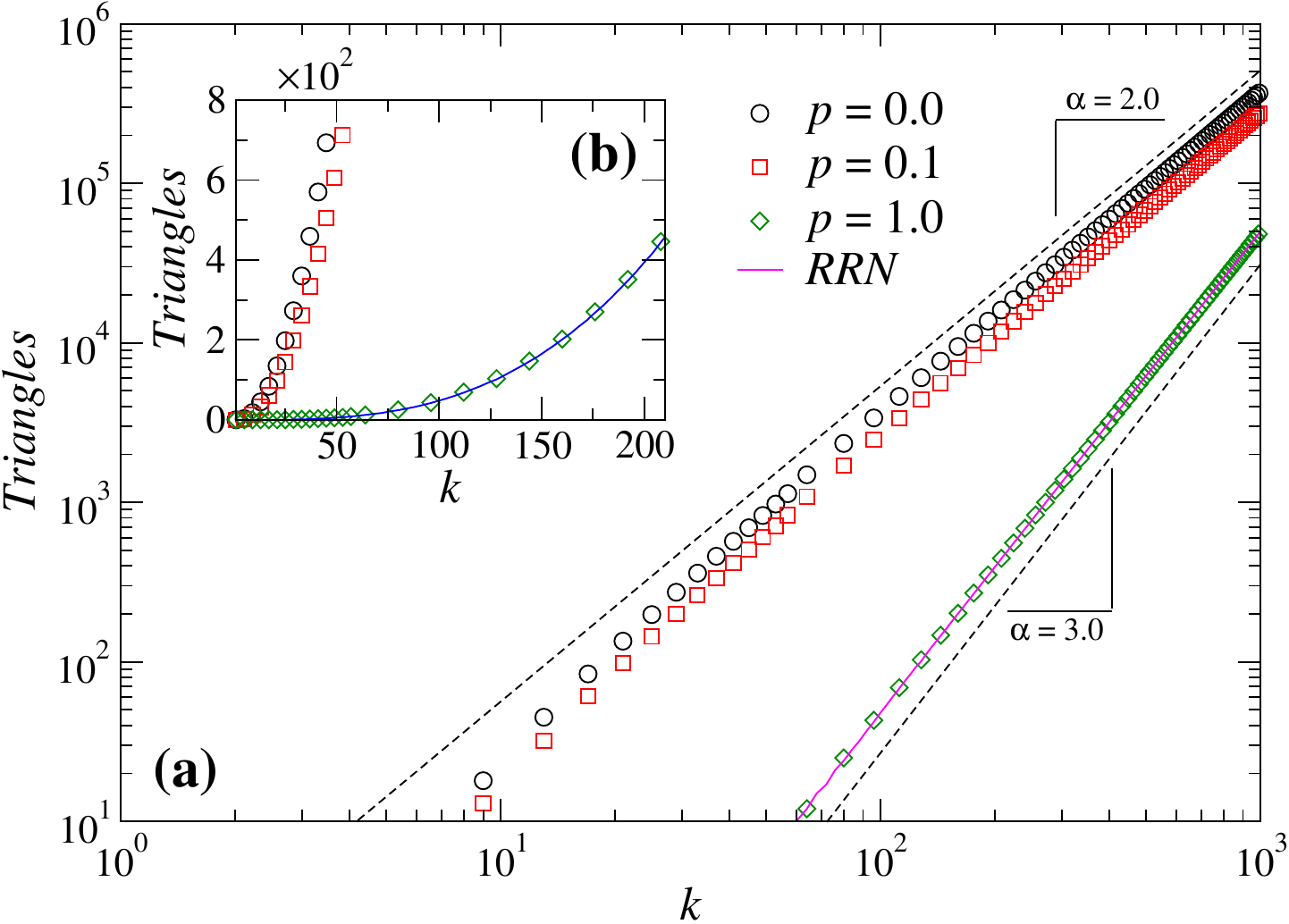}
 \caption{\textbf{\emph{Triangles}}. Panel $\mathbf{(a)}$: For curves $p=0.0$ and $p=0.1$, the lower the values of $k$, the greater the deviation from the analytical description - equation (\ref{eq.trn}) - until the curves assume a power-law description $f(k)\propto k^{\alpha}$ for higher values of the coordination number. As observed in figure \ref{fig01}, the curve for $p=1.0$, followed by that for RRN, one expects no corrections due to the strong presence of shortcuts on the WS model. It used WS networks with $N=10^4$ nodes.}
 \label{fig02}
\end{figure}

At the panel $\mathbf{(a)}$ of figure \ref{fig01} is shown the behavior of the transitivity and at figure \ref{fig02}, panel $\mathbf{(b)}$, the mean number of triangles, both when placed as a function of the coordination number $k$, for three representative values of the rewiring probability, namely $p=0.0$, $p=0.1$, and $p=1.0$. Unlike what is observed for $p=1.0$, regular and small-world scenarios display a fast increase of the transitivity up to values $k\approx24$. After this range, we attended for a decreasing growth rate, which can be noticed at the panel $\mathbf{(b)}$ of figure \ref{fig01}, reaching a (almost) constant behavior from values around $k\gtrsim24$. The curve for the small-world scenario has been observed to shift below the regular one due to the very characteristic of this structure; it is marked by the mean shortest path (chemical distance) among nodes, interpolating features from the regular structure and the random one \cite{watts1998collective} - indeed, there is a memory factor that comes from the original regular structure, as discussed in Ref. \cite{barrat2000properties}. The situation is different for the random scenario, for which we have not observed any substantial growth rate, with the transitivity curve smoothly increasing as the coordination number gets higher values.

Figure \ref{fig02}$\mathbf{(a)}$ shows the mean number of triangles, both regular and small-world scenarios, presenting a very similar behavior relying on a power-law description following $f(k)\propto k^{2}$ for values of $k>>1$. For lower values of $k$, the growth rate of triangles is greater, following $f(k)\propto k^{\alpha>2}$. On the other hand, no evidence was found on changing the power-law description for the random scenario. It assumes a logarithmic growth rate as the two different scenarios but scales faster with $\alpha=3$. For the sake of comparison, the panel $\mathbf{(a)}$ of both figures, \ref{fig01} and \ref{fig02}, show the mean values of transitivity and triangles, respectively, of an RRN structure. As expected, these curves follow those for the random scenario, $p=1.0$.

\subsection{Dynamical analysis}
We have investigated the evolutionary dynamics of PD and SG games concerning the behavior of the density of cooperators as a function of the temptation reward $T$ on small-world networks. Three scenarios of this network model were investigated according to the rewiring probability: $p=0.0$ (regular), $p=0.1$ (small-world), and $p=1.0$ (random). Mainly, it studied the coordination number $k$ range in which the transitivity property presents an increasing regime as a function of the nodes' degree. This function $\tau(k)$ assumes an almost constant behavior for values around $k=24$, as shown in figure \ref{fig01}. Equivalently, it will show investigations using $k=4$ and $k=16$, considering the deviation from the analytical point of view offered by equation (\ref{eq.trn}).

\begin{figure}[h]
 \centering
 \includegraphics[scale=0.33]{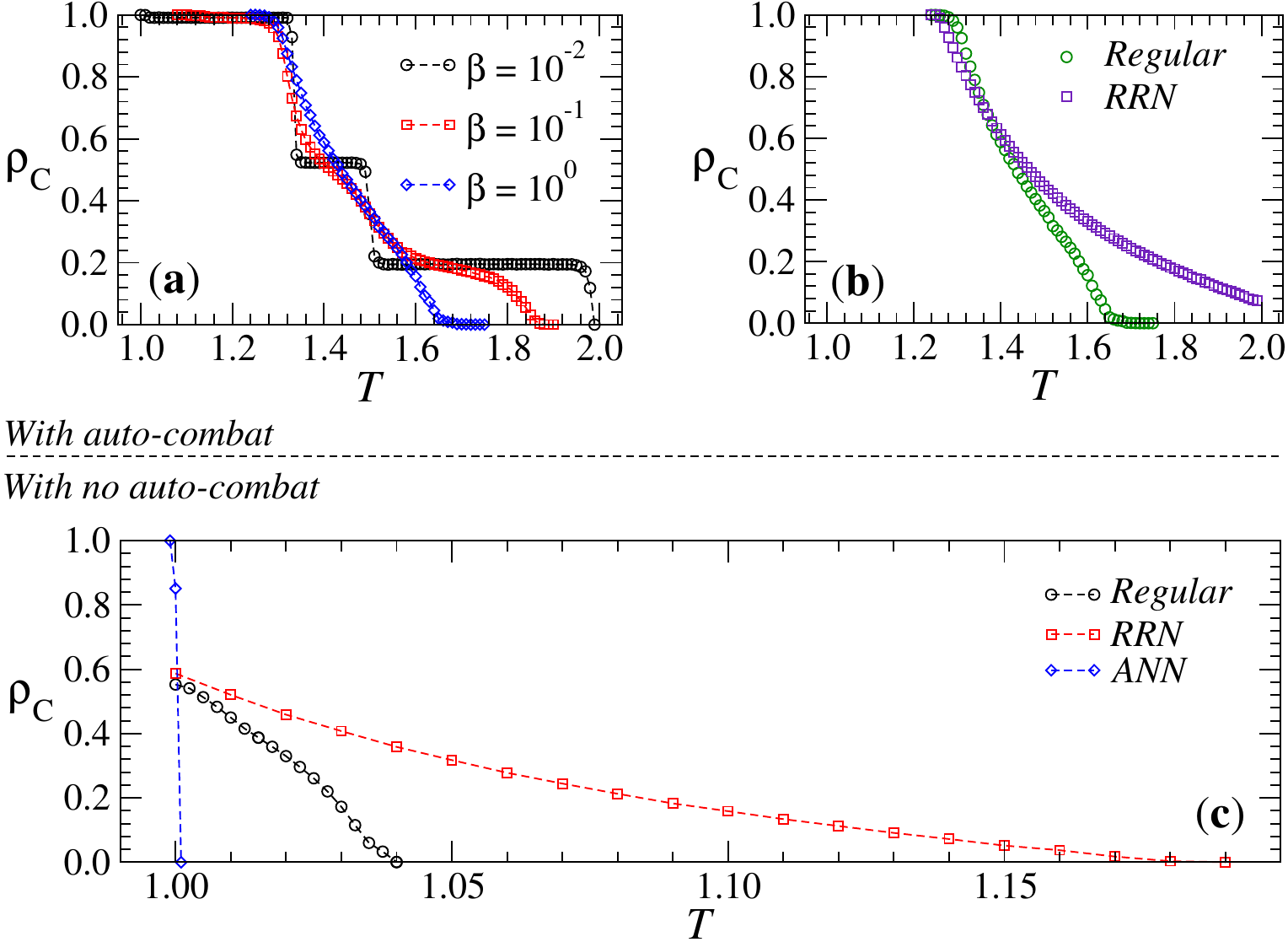}
 \caption{\textbf{\emph{Prisoner's Dilemma with and without auto-combat.}} At \textbf{(a)} and \textbf{(b)} are shown the PD dynamics with auto-combat. At \textbf{(a)}, the influence of the $\beta$ 	parameter over the promotion of cooperation is shown, whereas at \textbf{(b)} is shown the influence of the randomness, fixed $\beta=1.0$ and $k=4$. On the other hand, in panel \textbf{(c)}, the PD dynamics is shown with no auto-combat. The density of cooperators, as a function of the parameter $T$, was performed on different structures but with the same coordination number $k = 4$. Again, the role of randomness is shown but is also displayed in a comparison with annealed networks. Dynamics were performed on WS networks of size of $N=10^6$ nodes.} \label{fig03}
\end{figure}

Equivalently, it will show investigations using $k=4$ and $k=16$, considering the deviation from the analytical point of view offered by equation (\ref{eq.trn}). First, we have fixed the \emph{Sucker} parameter, setting $S=0.0$ and studied the influence of three features: (\textit{i}) The PD's dynamics allowing or not auto-combats, \emph{i.e.}, when in addition to playing with their neighbors, a player also plays with himself; (\textit{ii}) the role of perturbations on the suggestion process, allowing a strategy update. It is a kind of noise or, in physics meanings, the inverse of temperature; (\textit{iii}) the influence of randomness on the games' dynamics. It was done by using Regular, RRN, and ANN substrates. The results obtained from this first analysis are shown in figure \ref{fig03} and connect this work with others previously reported - For a more general review, readers can find on Ref. \cite{szabo2007evolutionary}. Various previous works have reported investigations using the auto-combat approach, which is shown in the \textbf{(a)} and \textbf{(b)} panels. Yet from \textbf{(a)} panel, it shows the role of the noise parameter $\beta$ on the PD's dynamics. To perform our analysis, we simulated the PD's dynamics with no auto-combat, and the noise parameter was set to $\beta=1.0$. It is shown on panel \textbf{(c)} of the exact figure. The comparison between \textbf{(b)} and \textbf{(c)} suggests that the absence of auto-combat makes the game more severe, decreasing the region where pure cooperative behavior takes place. Even in scenarios where a mix of strategies appears, cooperators and defectors fighting for their strategy, the survival of cooperators is reduced.

\begin{figure}[h]
 \centering
 \includegraphics[scale=0.33]{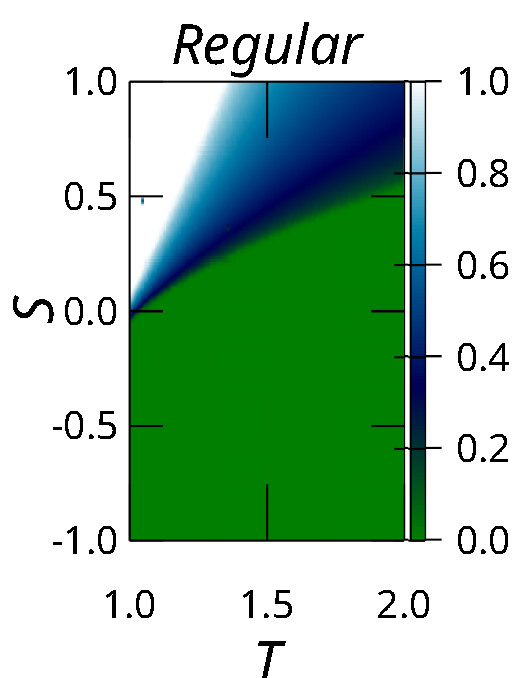}\includegraphics[scale=0.33]{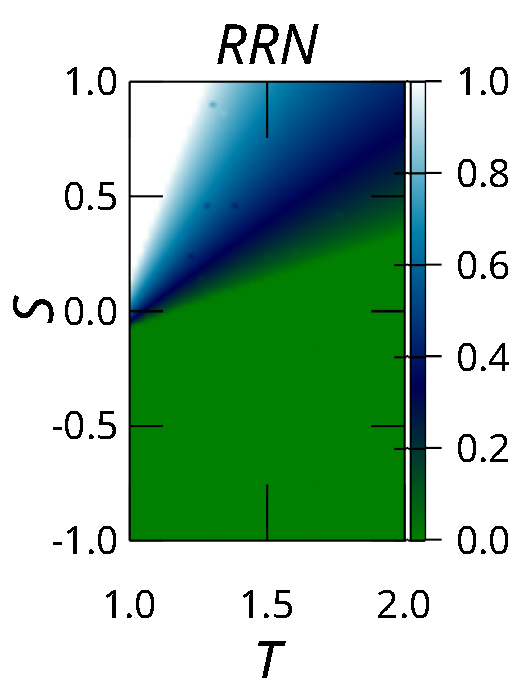}\includegraphics[scale=0.33]{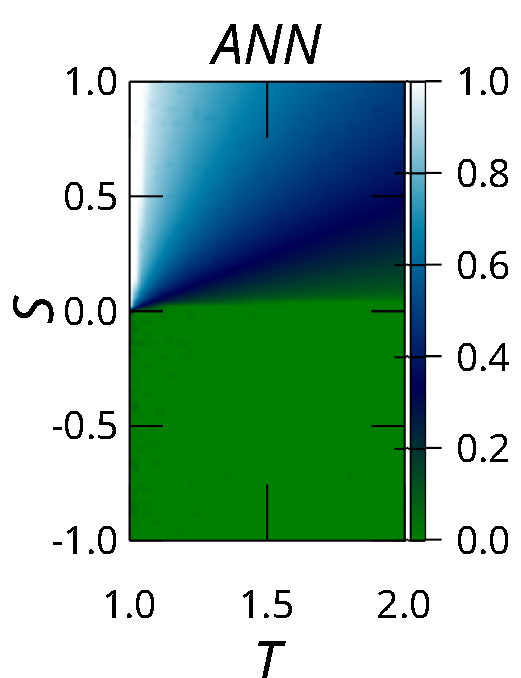}
 \caption{\textbf{\emph{Sucker versus Temptation.}} Density of cooperators on a crossed space Sucker \emph{vs.} Temptation. At each panel, colors correspond to the density of cooperators, which, in turn, is quantified on the right-hand side scale. It shows (from the left to the right) the following structures: (a) Regular Networks, (b) Random Regular Networks, and (c) Annealed Networks. These patterns were used to compare those found for small-world networks. The parameters used were $k=4$ and $N=10^4$. From the left to the right, one sees the increase of the mixture pattern (cooperators and defectors) in the Snowdrift Game scale, $S:[0,1]$, as the parameter $T$ increases. On the other hand, the PD's dynamics, $S:[-1,0]$, becomes more severe as on Regular and ANN networks, with the cooperative behavior reduced to a small region close to $T = 1$ and $S = 0$. This latter is also reached by the solution of the replicator equation (\ref{eq1}).}
 \label{fig04}
\end{figure}

To visualize the domains of the cooperative, defecting, and coexistence of strategies, figure \ref{fig04} shows heat maps crossing parameters $S$ and $T$. From the left to the right, a colored pallet for the density scale of cooperators is shown when the evolutionary games occurred on top of Regular, RRN, and ANN. It was analyzed $T:[1,2]$ and $S:[-1,0]$ for PD's dynamics and $S:[0,1]$ concerning the same $T$ interval, corresponding to the SG's dynamics. Since ANN is recognized to describe systems on a mean-field approach, what is shown in figure \ref{fig04}  (from left to right) is the decreasing of degree-degree correlations. This decrease affects both games; PD's dynamics become more severe for Regular and ANN networks (close to $T=1.0$), and SG's dynamics display an increasing region of strategies coexistence since SG players share their costs. The randomness introduced by the RRN model is relevant to promoting cooperation; both Regular and RRN were constructed with $k=4$. The difference appears in how connections were made: Regular, in the sense of a metric, and RRN with paring made by chance. For ANN networks, the time interval used to reconnect pairs of nodes was smaller than the characteristic time scale of the dynamics.

\begin{figure}[h]
 \centering
 \includegraphics[width=0.48\textwidth]{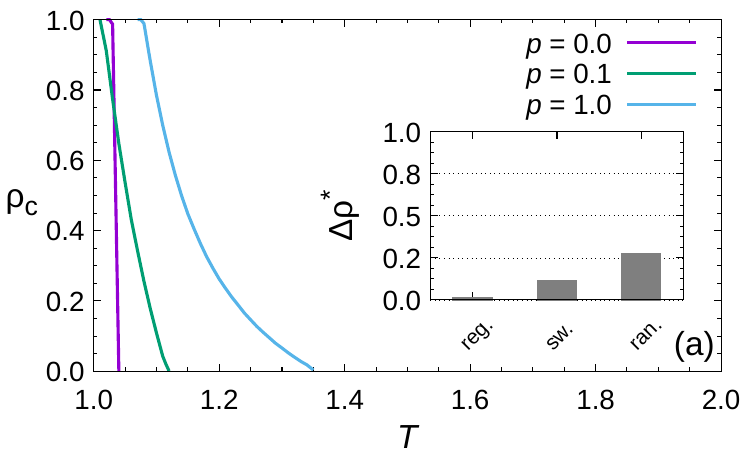}
 \includegraphics[width=0.48\textwidth]{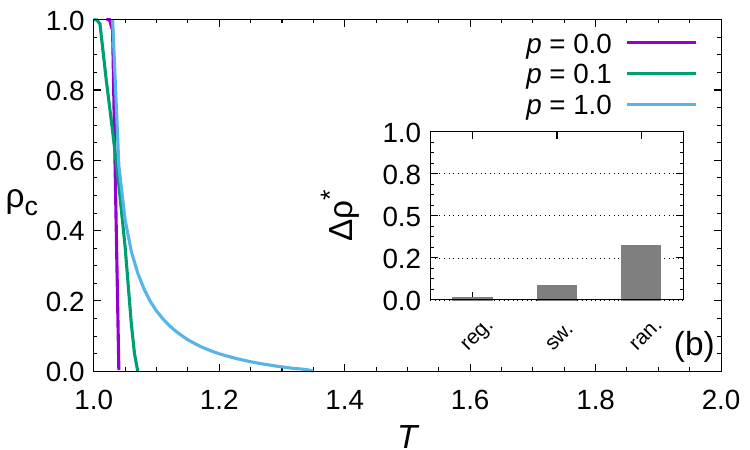}
 \caption{\textbf{\emph{Small-world topology and the cooperative behavior.}} The density of cooperators is shown as a function of the temptation reward when the evolutionary game is played on WS networks. The Sucker reward was set to $S=0.0$. Three representative values for the rewiring probability were used, namely, $p=0.0$ (Regular), $p=0.1$ (Small-World), and $p=1.0$ (Random). Panel (a) used networks with the coordination number $k=4$; at (b), this value was $k=16$. Both analysis was performed using networks of size of $N=10^6$ nodes.}
 \label{fig05}
\end{figure}

Figure \ref{fig05} shows the density of cooperators $\rho_c$ as a function of the temptation reward $T$, on top of small-world networks, with the parameter $S=0.0$. Three values of the rewiring probability $p$ were analyzed, considering representative ensembles corresponding to three different topologies: Regular, Small-World, and Random. As $p$ goes from zero to one (equivalently, as the randomness feature is incremented), the region of strategies coexistence also increases. We have compared the size of this region for each value of $p$, estimating values for ordered-disordered transition points. So, denoting $\rho_c^*$ as that one corresponding to the fully ordered cooperative phase to a coexistence of cooperators and defectors, and denote $\rho_d^*$ as the transition point from that of the coexistence of strategies region to the fully ordered defective behavior. Thus, $\Delta\rho_k^*=(\rho_d^*-\rho_c^*)$ is the estimated size of the region of strategies coexistence, where cooperators and defectors coexist. It is shown on the insets of figure \ref{fig05}. Interestingly, considering $p>0.0$, we have observed $\Delta\rho_{16}^*>\Delta\rho_4^*$, although it is due to discouragement of the cooperative behavior, since $\rho_{c,k=16}^*<\rho_{c,k=4}^*$. For a more detailed discussion on transition points concerning the PD game on a small-world model, please see Ref. \cite{PhysRevE.77.026109}.

\begin{figure}[h]
        \centering
        \includegraphics[scale=0.20]{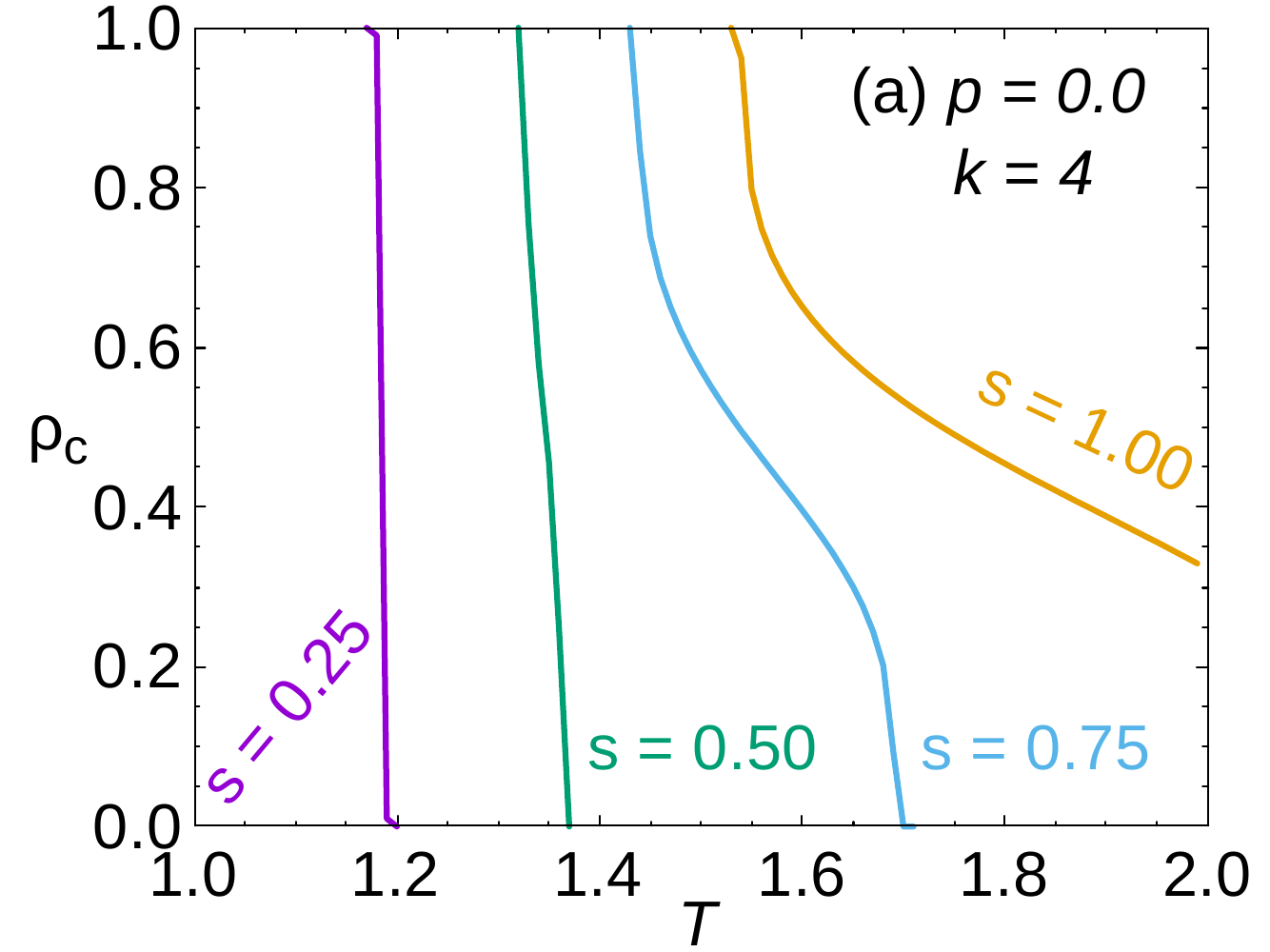}\includegraphics[scale=0.20]{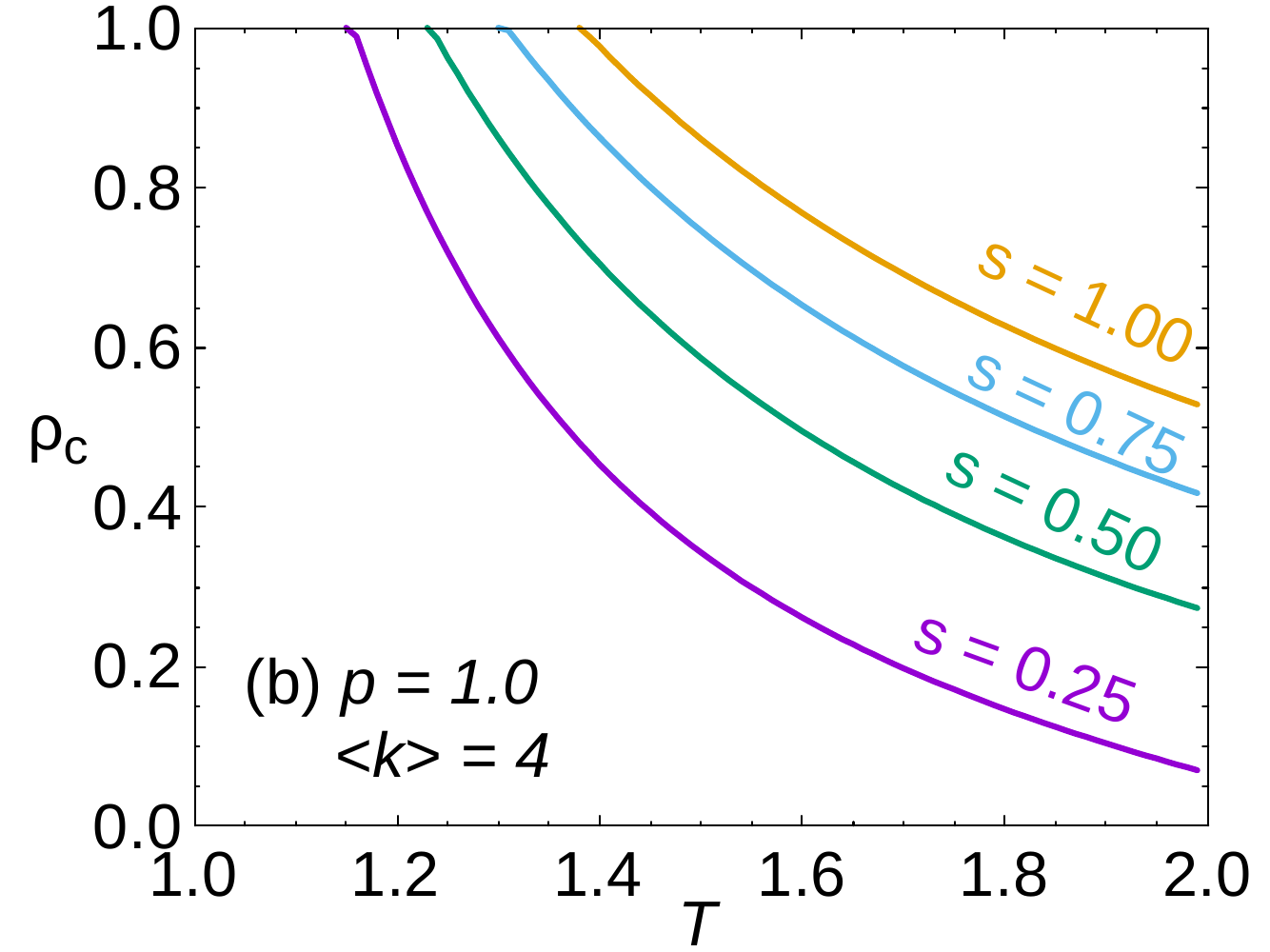}\\
        \includegraphics[scale=0.20]{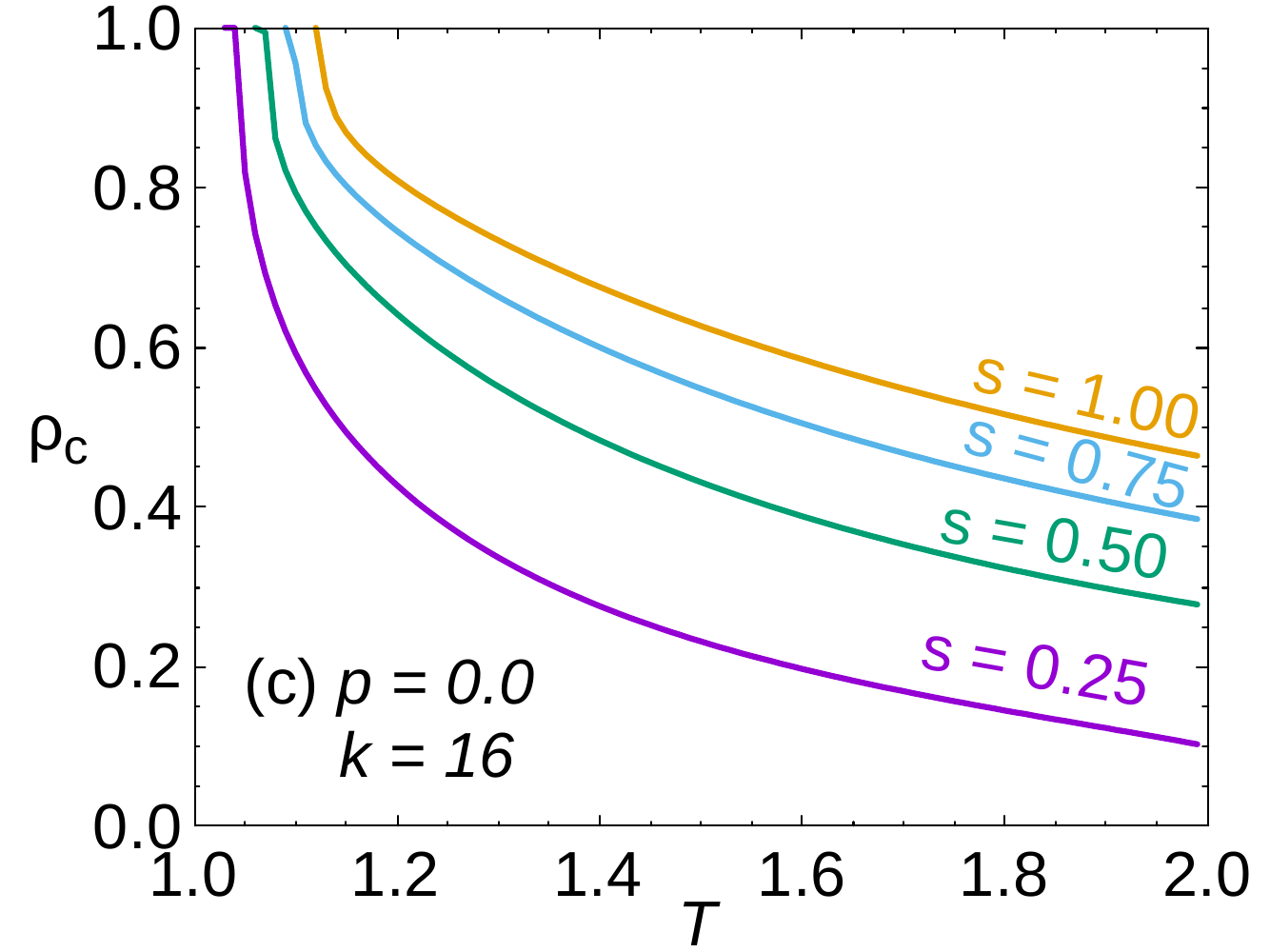}\includegraphics[scale=0.20]{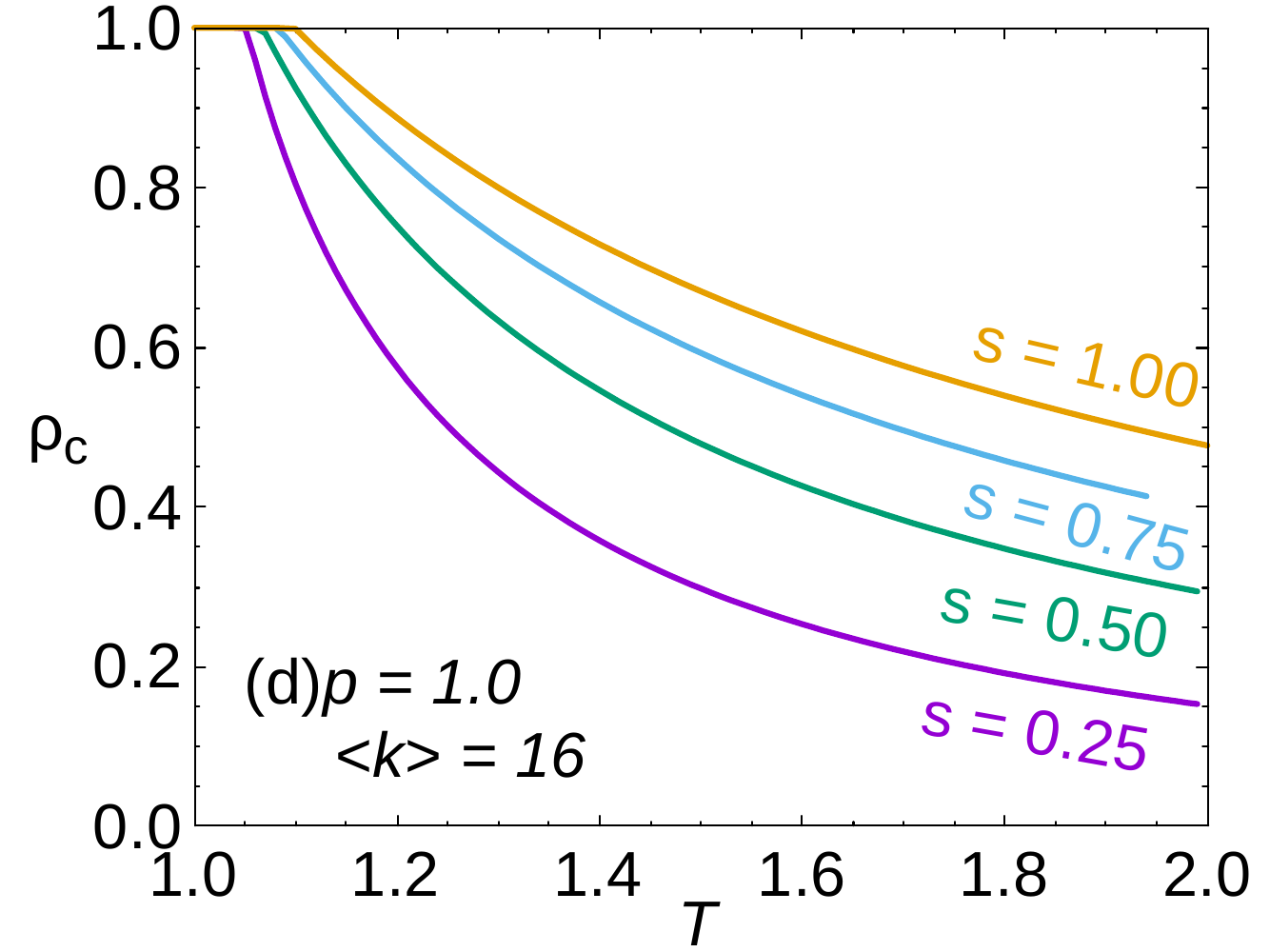}
            \caption{\textbf{Snowdrift Game.} Panels (a)-(d) show the curves for the density of cooperators as a function of the temptation, $T$. At (a), the network is $k-regular$, in which all nodes have $k=4$ neighbors. The same for panel (c), but $k=16$. At (b), the ensemble network has, on average, $\langle k\rangle = 4$, whereas at panels (d) $\langle k\rangle = 16$. On the left column, $p = 0.0$, and on the right, $p=1.0$. The coexistence of strategy patterns increases its region as the WS network goes from a regular to a random structure. On the other hand, pure cooperation is affected by this increase. These analyses were performed using networks of size of $n = 10^6$ nodes.}
        \label{fig06}
    \end{figure}

For values of the parameter $S:[0,1]$, the curves of the density of cooperators as functions of the temptation $T$ are shown in figure \ref{fig06}. Panels (a) and (b) correspond to WS networks with $p=0.0$ and $p=1.0$, respectively. Both structures were prepared with $k=4$, which, in turn, for the random scenario ($p=1.0$), means that $\langle k\rangle=k$. The same analysis is shown in panels (c) and (d), although it was used $k=16$. Since the SG is a more cooperative game with shared costs, the region of strategy coexistence increases with $S$. Moreover, the randomness ingredient promotes a region of coexistence of strategies, leading to a lack of the transition from this scenario to a pure defecting behavior. Only for values of $p$ close to zero and low values of the degree $k$, we observe the transition point $\rho_d^*$. It was only observed in panel (a) for cases where $S<1$. On the other hand, this is accompanied by a retreat of the value of the transition point $\rho_c^*$. It gets clearer by observing those values of $k$ in which the transitivity assumes a constant regime as a function of $k$ - see figure \ref{fig01}, as shown in panels (c) and (d).
    \begin{figure*}[htb]
        \centering
        \includegraphics[scale=.33]{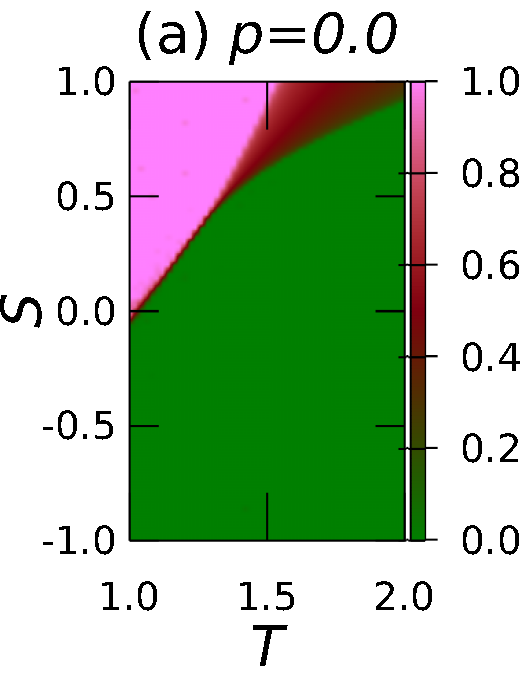}~~\includegraphics[scale=.33]{fig6b2.pdf}~~\includegraphics[scale=.33]{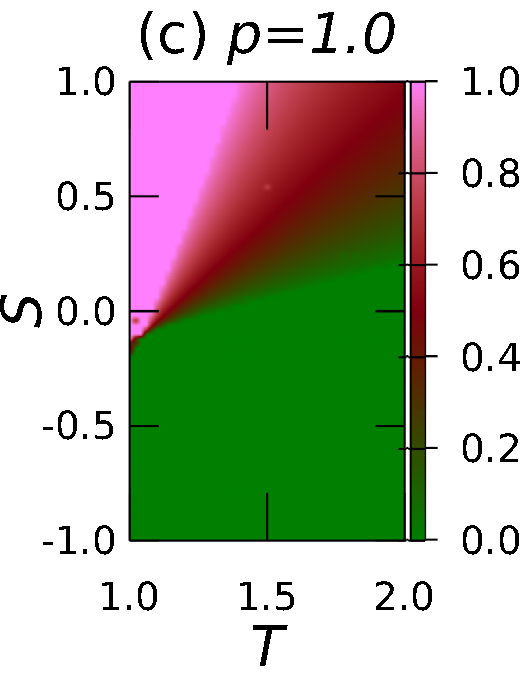}
        \includegraphics[scale=.335]{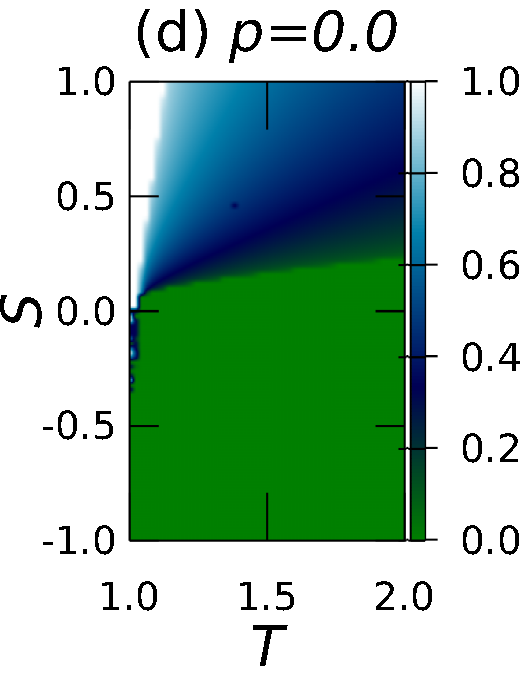}~~\includegraphics[scale=.335]{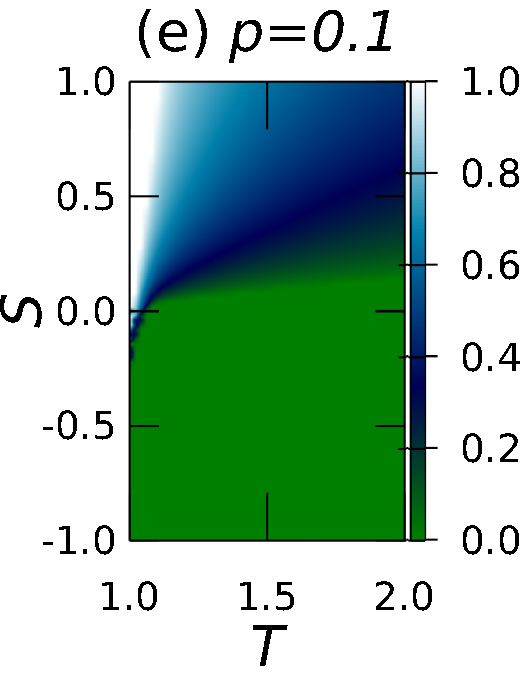}~~\includegraphics[scale=.335]{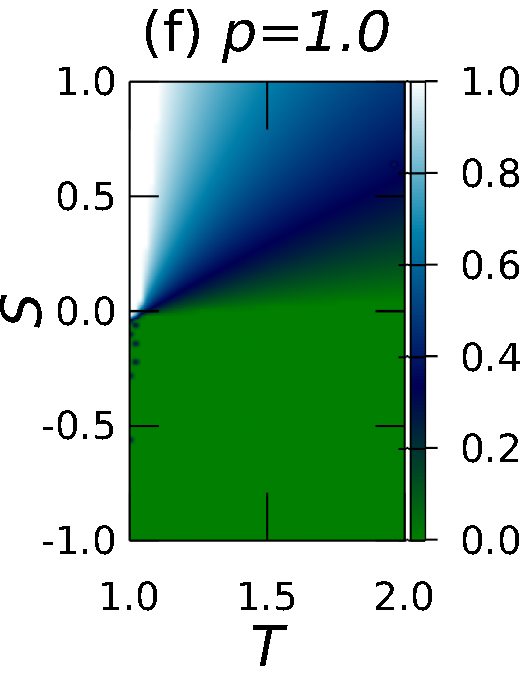}
            \caption{\textbf{Sucker \emph{versus} Temptation.} Density of cooperators on a crossed space Sucker \emph{vs.} Temptation. The evolutionary dynamics were performed on top of the WS model with $N=10^4$ and different rewiring probabilities: $p=0.0$ (Regular), $p=0.1$ (Small-World), and $p=1.0$ (Random). From panel (a) to (c), the values used for the coordination number were $k=4$, whereas $k=16$ from (d) to (f).  In the scale of the Snowdrift Game, $S:(0,1]$, the mixture pattern (coexistence of strategies) increases as with the parameter $T$. On the other hand, the PD's dynamics, $S:[-1,0]$ becomes more severe likewise on ANN networks, with the pure cooperative behavior reduced to a small region close to $T = 1$.}
            \label{fig07}
    \end{figure*}

The information from previous figures concerning dynamical features can be compared using the following set of heat maps shown in figure \ref{fig07}. This figure studied PD's and SG's dynamics on top of WS networks. These dynamics was performed using $p=0.0$, $p=0.1$, and $p=1.0$ for networks with $k=4$, panels (a) to (c), and $k=16$ from panels (d) to (f). As a guide, these densities of cooperators' figures can be compared with those in figure \ref{fig03}. So, the model with $p=0.0$ and $p=0.1$ can be compared to the Regular structure - Figures 7(a) and 7(b) compared with figure \ref{fig04} Regular. Taking the increase of the coordination number of $k=4$ to $k=16$, the PD's dynamics become more severe, whereas SG's dynamics display an increase in the region of the coexistence of strategies. Moreover, when comparing the WS model with $p=1.0$ for RRN and ANN networks, the increase of the coordination number induces a decrease in the full cooperative behavior, close to $T=1.0$, for the PD game. On the other hand, the effect on SG's dynamics was that of increases in the coexistence of strategies, been the case of $\langle k\rangle=4$ compared with RRN and $\langle k\rangle=16$ compared with ANN networks - Figure 7(c) and 7(f) compared with figure \ref{fig04} RRN and \ref{fig04} ANN, respectively.

Three aspects can be taken into account concerning regions with (\emph{i}) pure cooperation, (\emph{ii}) pure defecting, and (\emph{iii}) coexistence of strategies. The region of pure cooperation, for PD dynamics, \emph{i.e.}, the square $T:[1,2]$ and $S:[-1,0]$, increases with the randomness feature of the structure. However, do not increase as the number of neighbors does from $k=4$ to $k=16$. It can be observed in the regions close to $T=1$. This game remains severe, even increasing the number of neighbors, with no substantial modifications for the pure defecting region and a small region with the coexistence of strategies on the border of transition between pure strategy regions.

This fact that increasing the number of neighbors does not favor the PD dynamics remains if considering the SG game, even recognized as a game with sharing costs. Considering the square $T:[1,2]$ and $S:(0,1]$, the pure cooperative behavior decreases as $k$ goes from value 4 to 16 neighbors. On the other hand, with the increase in the value of $k$, the coexistence of strategies is favored. Comparing with figure \ref{fig04}, as the transitivity $\tau(k)$ assumes a constant behavior, the stationary regime gets closer to the annealed pattern. It leads the dynamical process to the mean-field expectation.

\section{Conclusions}
\label{sec:conc}
We have explored the dynamics of Prisoner's Dilemma and Snowdrift Game, two evolutionary game models occurring on small-world networks. The cooperative behavior of these dynamics was investigated, assuming the network's transitivity was increasing. The transitivity property is related to the network reciprocity, which, in turn, is recognized as a rule of the evolution of cooperation \cite{nowak2006five}. This defense mechanism of cooperative behavior is based on the fact that cooperators can avoid defectors by forming clusters.

For structural analysis, we have shown that the regime of an increasing network reciprocity is marked by a deviation between analytical and simulation solutions of the transitivity property for the regular and small-world structures from the Watts-Strogatz model \cite{watts1998collective,barrat2000properties}. This discrepancy occurs for low coordination number values, up to $k\lesssim24$. In doing so, we restrict our analyses to the degrees $k=4$ and $k=16$, covering that increasing regime of the transitivity. As a function of the coordination number, the number of triangles is size-independent for random WS networks ($p=1.0$) and random regular networks, ruled by $\sim k^{\alpha=3.0}$. However, for regular and small-world structures, this function is $\propto k^{\alpha>2.0}$, in the regime studied, up to $k\lesssim20$. For values even greater than $k$, this function assumes an asymptotic scale $\sim k^{\alpha=2.0}$. The range $2<\alpha<3$ is characterized by an increase in shortcuts due to the mechanism of link rewiring. Therefore, the average short path length among nodes decreases using the arrangement of triangles.

For dynamical analysis in the increasing regime of the transitivity property, in general terms, the Snowdrift Game converges to an annealed scenario as randomness and coordination numbers increase. In contrast, the Prisoner's Dilemma becomes more severe against the cooperative behavior. Comparing WS-topologies $p=0.0$ and $p=1.0$ with Regular networks, the network reciprocity increases, particularly from $k=4$ to $k=16$.  The coexistence of strategies is promoted under the scale of the Snowdrift Game, whereas PD's dynamics show no substantial changes. However, the rewiring process ($p=0.1$) promotes cooperation when $T$ is close to one.

On the other hand, in the comparison of WS $p=1.0$ (Random scenario) to RRN and ANN networks, the regime of increasing network reciprocity, the Snowdrift Game dynamics goes from the pattern of an RRN network to an ANN one, as the coordination number is changed from $k=4$ to $k=16$. It corresponds to the mean-field scenario. Yet, in the region of coexistence of strategy, meaning the distance between transition points of pure cooperative behavior and pure defecting one, this region is greater for $k=16$ than over $k=4$. However, it is due to discouragement of the cooperative behavior since the pure cooperative transition point for $k=16$ takes place before that one for $k=4$.

Our work may stimulate further in-depth investigation, particularly on three aspects: First, from a structural point of view, a question remains open concerning the deviation between analytical and simulation solutions of the transitivity property on other small-world structures, different from Watts-Strogatz networks. For instance, those reported in Ref. \cite{PhysRevE.77.026109}, for which there is not a fixed density of random ``shortcuts'' on the lattice, connecting nodes for distant regions and in Ref. \cite{Gang2008} which, in turn, the small-world effect appears combined with an unchanged connectivity for each node. Second, on the impact of the $\beta$ parameter (temperature-like) on the evolution of cooperation in the presence of auto-combat. Last, investigations of critical properties in an increasing network reciprocity regime combine size-dependence and scaling relations to determine critical exponents when these systems undergo phase transitions and other types of spreading models, such as synchronization and opinion dynamics.

\section{Acknowledgments}
F.B.P. would like to thank Universidade Federal de Ouro Preto (UFOP) under project 19/2019-PIP-2S/UFOP-2019-20; R.S.F. would like to thank Fundação de Amparo à Pesquisa do Estado de Minas Gerais (FAPEMIG) for the financial support;  D.S.M.A., T.F.A.A, G.A.A., F.W.S.L., and A.M-F. would like to thank Coordenação de Aperfeiçoamento de Pessoal de Nível Superior (CAPES), Conselho Nacional de Desenvolvimento Científico e Tecnológico (CNPq), Fundação de Amparo a Pesquisa do Estado do Piauí (FAPEPI) for the financial support.





\begin{thebibliography}{99}

\bibitem{hofbauer2003evolutionary}
J. Hofbauer and K. Sigmund,
\newblock Evolutionary game dynamics,
\newblock Bull. Amer. Math. Soc. 40 (2003) 479-519.

\bibitem{axelrod1981evolution}
R. Axelrod and W.D. Hamilton,
\newblock The evolution of cooperation,
\newblock Science 211 (1981) 1390-1396.
\newblock https://doi.org/10.1126/science.7466396.

\bibitem{turner1999prisoner}
P.E. Turner and L. Chao,
\newblock Prisoner's dilemma in an RNA virus,
\newblock Nature 398 (1999) 441-443.
\newblock https://doi.org/10.1038/18913.

\bibitem{szabo2007evolutionary}
G. Szabó and G. Fath,
\newblock Evolutionary games on graphs,
\newblock Phys. Rep. 446 (2007) 97-216.
\newblock https://doi.org/10.1016/j.physrep.2007.04.004.

\bibitem{santos2014biased}
M.D. Santos, S.N. Dorogovtsev, J.F.F. Mendes, 2014.
\newblock Biased imitation in coupled evolutionary games in interdependent networks.
\newblock Sci. Rep. 4, 04436.
\newblock https://doi.org/10.1038/srep04436.

\bibitem{nowak1992evolutionary}
M.A. Nowak and R.M. May,
\newblock Evolutionary games and spatial chaos,
\newblock Nature 359 (1992) 826-829.
\newblock https://doi.org/10.1038/359826a0.

\bibitem{hauert2004spatial}
C. Hauert and M. Doebeli,
\newblock Spatial structure often inhibits the evolution of cooperation in the
  snowdrift game,
\newblock Nature 428 (2004) 643-646.
\newblock https://doi.org/10.1038/nature02360.

\bibitem{nowak2006five}
M.A. Nowak,
\newblock Five rules for the evolution of cooperation,
\newblock Science 314 (2006) 1560-1563.
\newblock https://doi.org/10.1126/science.113375.

\bibitem{roca2009effect}
C.P. Roca, J.A. Cuesta, A. S{\'a}nchez, 2009.
\newblock Effect of spatial structure on the evolution of cooperation,
\newblock Phys. Rev. E. 80, 046106.
\newblock https://doi.org/10.1103/PhysRevE.80.046106.

\bibitem{PhysRevE.77.026109}
J. Vukov, G. Szab\'o, A. Szolnoki, 2008.
\newblock Evolutionary prisoner's dilemma game on Newman-Watts networks.
\newblock Phys. Rev. E. 77, 026109.
\newblock https://doi.org/10.1103/PhysRevE.77.026109.

\bibitem{PhysRevE.89.052813}
Z. Wang, L. Wang, M. Perc, 2014.
\newblock Degree mixing in multilayer networks impedes the evolution of cooperation.
\newblock Phys. Rev. E. 89, 052813.
\newblock https://link.aps.org/doi/10.1103/PhysRevE.89.052813.

\bibitem{PhysRevE.82.047101}
Z. Rong, H-X. Yang, W-X. Wang, 2010.
\newblock Feedback reciprocity mechanism promotes the cooperation of highly clustered scale-free networks.
\newblock Phys. Rev. E. 82, 047101.
\newblock https://link.aps.org/doi/10.1103/PhysRevE.82.047101.

\bibitem{PhysRevE.78.066101}
M. Perc, A. Szolnoki, G. Szab\'o, 2008.
\newblock Restricted connections among distinguished players support cooperation.
\newblock Phys. Rev. E. 78, 066101.
\newblock https://link.aps.org/doi/10.1103/PhysRevE.78.066101.

\bibitem{chen2008interaction}
X. Chen, F. Fu, L. Wang, 2008.
\newblock Interaction stochasticity supports cooperation in spatial prisoner’s dilemma
\newblock Phys. Rev. E. 78, 051120.
\newblock https://doi.org/10.1103/PhysRevE.78.051120.

\bibitem{PhysRevE.80.056112}
A. Szolnoki, J. Vukov, G. Szab\'o, 2009.
\newblock Selection of noise level in strategy adoption for spatial social dilemmas.
\newblock Phys. Rev. E. 80, 056112.
\newblock https://doi.org/10.1103/PhysRevE.80.056112.

\bibitem{Perc_2011}
M. Perc, 2011.
\newblock Does strong heterogeneity promote cooperation by group interactions?
\newblock New J. Phys. 13, 123027.
\newblock https://dx.doi.org/10.1088/1367-2630/13/12/123027.

\bibitem{JIA2024115333}
C-X. Jia, L. Ma, R-R. Liu, 2024.
\newblock Enhancing cooperation through payoff-related inertia in networked prisoner’s dilemma game.
\newblock Chaos Solit. Fractals. 186, 115333.
\newblock https://doi.org/10.1016/j.chaos.2024.115333.

\bibitem{Li_2024}
P. Li, T. Ye, S. Fan, 2024.
\newblock The impact of preferential selection mechanism based on average payoff and stability of payoff on the evolution of cooperation.
\newblock Europhys. Lett. 146, 21003.
\newblock https://dx.doi.org/10.1209/0295-5075/ad3188.

\bibitem{A_li2016}
A. Li, M. Broom, J. Du, L. Wang, 2016.
\newblock Evolutionary dynamics of general group interactions in structured populations.
\newblock Phys. Rev. E. 93, 022407.
\newblock https://link.aps.org/doi/10.1103/PhysRevE.93.022407.

\bibitem{PhysRevE.79.067101}
S. Meloni, A. Buscarino, L. Fortuna, M. Frasca, J. G\'omez-Garde\~nes, V. Latora, Y. Moreno, 2009.
\newblock Effects of mobility in a population of prisoner's dilemma players.
\newblock Phys. Rev. E. 79, 067101.
\newblock https://link.aps.org/doi/10.1103/PhysRevE.79.067101.

\bibitem{Wu2017}
Y. Wu, S. Shuhua, Z. Zhang, Z. Deng, 2017.
\newblock Impact of Social Reward on the Evolution of the Cooperation Behavior in Complex Networks.
\newblock Sci. Rep. 7, 41076.
\newblock https://doi.org/10.1038/srep41076.

\bibitem{Gang2008}
W. Gang, G. Kun, Y. Han-Xin and W. Bing-Hong, 2008.
\newblock Role of Clustering Coefficient on Cooperation Dynamics in Homogeneous Networks.
\newblock Chinese Phys. Lett. 25 2307.
\newblock https://doi.org/10.1088/0256-307X/25/6/105.

\bibitem{taylor1978evolutionary}
P.D. Taylor and L.B. Jonker,
\newblock Evolutionary stable strategies and game dynamics,
\newblock Math. Biosci. 40 (1978) 145-156.
\newblock https://doi.org/10.1016/0025-5564(78)90077-9.

\bibitem{watts1998collective}
D.J. Watts and S.H. Strogatz,
\newblock Collective dynamics of ``small-world'' networks,
\newblock Nature 393 (1998) 440-442.
\newblock https://doi.org/10.1038/30918.












\bibitem{albert2002statistical}
R. Albert and A-L. Barab{\'a}si,
\newblock Statistical mechanics of complex networks,
\newblock Rev. Mod. Phys. 74 (2002) 47-97.
\newblock https://doi.org/10.1103/RevModPhys.74.47.

\bibitem{barrat2000properties}
A. Barrat and M. Weigt,
\newblock On the properties of small-world network models.
\newblock Eur. Phys. J. B. 13 (2000) 547-560.
\newblock https://doi.org/10.1007/s100510050067  
  
\bibitem{dorogovtsev2008critical}
S.N. Dorogovtsev, A.V. Goltsev, J.F.F. Mendes,
\newblock Critical phenomena in complex networks,
\newblock Rev. Mod. Phys., 80 (2008) 1275-1335.
\newblock https://doi.org/10.1103/RevModPhys.80.1275

\bibitem{dorogovtsev2010lectures}
S. Dorogovtsev,
\newblock Lectures on complex networks,
\newblock OUP, 2010.
\newblock https://doi.org/10.1093/acprof:oso/9780199548927.001.0001

\bibitem{Ferreira2013}
R.S. Ferreira, S.C. Ferreira, 2013.
\newblock Critical behavior of the contact process on small-world networks.
\newblock Eur. Phys. J. B. 86, 462.
\newblock https://doi.org/10.1140/epjb/e2013-40534-0

\bibitem{PhysRevE.83.066113}
S.C. Ferreira, R.S. Ferreira, R. Pastor-Satorras, 2011. 
\newblock Quasistationary analysis of the contact process on annealed
  scale-free networks.
\newblock Phys. Rev. E. 83, 066113.
\newblock https://doi.org/10.1103/PhysRevE.83.066113.

\bibitem{luce1949method}
R.D. Luce and A.D. Perry,
\newblock A method of matrix analysis of group structure,
\newblock Psychometrika, 14 (1949) 95-116.
\newblock https://doi.org/10.1007/BF02289146.

\bibitem{wasserman1994social}
S. Wasserman and K. Faust,
\newblock Social network analysis: Methods and applications, vol.~8,
\newblock CUP, 1994.
\newblock https://doi.org/10.1017/CBO9780511815478.

\bibitem{milo2002network}
R. Milo, S. Shen-Orr, S. Itzkovitz, N. Kashtan, D. Chklovskii, U. Alon,
\newblock Network motifs: Simple building blocks of complex networks,
\newblock Science 298 (2002) 824-827.
\newblock https://doi.org/10.1126/science.298.5594.824.

\end{thebibliography}

\end{document}